# Localization of Lorentz transformation and its induced local Lorentz invariance

Jian-Miin Liu*
Department of Physics, Nanjing University
Nanjing, The People's Republic of China
*On leave. E-mail address: liu@phys.uri.edu, miinliu@hotmail.com

ABSTRACT
Introducing the primed inertial coordinate system, for each inertial frame of reference, in addition to the usual inertial coordinate system, we assume that gravity-free space and time possess the Euclidean structures in the primed inertial coordinate system and the generalized Finslerian structures in the usual inertial coordinate system. We combine these assumptions with two fundamental postulates, (i) the principle of relativity and (ii) the constancy of the speed of light in all inertial frames of reference, to derive the localized Lorentz transformation as a linear transformation between any two usual inertial coordinate systems. Based on this, it is proposed that all laws of physics are locally Lorentz-invariant in the usual inertial coordinate system. As a Lorentz-invariant law of physics must be locally Lorentz-invariant while a locally Lorentz-invariant law is not necessarily Lorentz-invariant, the change from the requirement of Lorentz invariance to that of the local Lorentz invariance on laws of physics provides with a larger acceptable scope to explore these laws. The localization of Lorentz transformation and its induced local Lorentz invariance precisely meet the demands of Einstein's theory of gravitation.

## 1. INTRODUCTION

Lorentz was the first who learned that transformation between two usual inertial coordinate systems, $\{x_m^r, t_m\}$, $r=1,2,3$, separately belonging to two inertial frames of reference m (IFRm), m=1,2,

$$x_2^r = (x_1^r - u^r t_1) + (\gamma-1)u^r u^s (x_1^s - u^s t_1)/u^2, \quad r,s=1,2,3, \tag{1a}$$
$$t_2 = \gamma(t_1 - u^k x_1^k/c^2), \quad k=1,2,3, \tag{1b}$$

leaves Maxwell's equations of electromagnetic fields invariant [1], where $u^s$, s=1,2,3, is the relative velocity of IFR2 to IFR1, $u=(u^s u^s)^{1/2}$ and

$$\gamma = 1/(1-u^2/c^2)^{1/2}. \tag{2}$$

The transformation now bears his name. Here, the usual inertial coordinate system in inertial frame of reference is in Einstein's definition which will be quoted below. Lorentz's this discovery made theoretical physics situated: Galilean-invariant mechanics and Lorentz-invariant (invariant under the Lorentz transformation) electrodynamics co-existed. A question was naturally raised: Of the Galilean transformation and the Lorentz transformation, which one is the actual transformation between the two usual inertial coordinate systems?

Believing in the principle of relativity and Maxwell's equations of electromagnetic fields, in his special theory of relativity, Einstein derived the Lorentz transformation as a linear transformation between any two usual inertial coordinate systems and proposed that all laws of physics in nature are Lorentz-invariant in the usual inertial coordinate system [2,3]. He also developed Lorentz-invariant mechanics for a particle [2,3]. Since, the Lorentz invariance has become a physical principle: In the usual inertial coordinate system, all physical laws keep their forms under the Lorentz transformation. This principle has been very successful in exploring laws of physics in various, though not all, fields.

In this paper, we derive the localized Lorentz transformation as a linear transformation between any two usual inertial coordinate systems and propose that all laws of physics are locally Lorentz-invariant (invariant under the localized Lorentz transformation) in the usual inertial coordinate system. The paper consists of seven sections: introduction, justification for localizing Lorentz transformation, derivation of Lorentz transformation, local structures of gravity-free space and time, localized Lorentz transformation, relativistic velocity space, conclusion and discussion.

## 2. JUSTIFICATION FOR LOCALIZING LORENTZ TRANSFORMATION

In contrast to Lorentz transformation, the localized Lorentz transformation between two usual inertial coordinate systems, $\{x_m^r, t_m\}$, r=1,2,3, of IFRm, m=1,2, is in the shape of

$$dx_2^r = (dx_1^r - u^r dt_1) + (\gamma-1)u^r u^s (dx_1^s - u^s dt_1)/u^2, \quad r,s=1,2,3, \tag{3a}$$

$$dt_2=\gamma(dt_1-u^k dx_1^k/c^2), \quad k=1,2,3, \tag{3b}$$

everywhere and every time.

Logically, the Lorentz transformation covers the localized Lorentz transformation and the localized Lorentz transformation does not necessarily lead to Lorentz transformation. So, a Lorentz-invariant law must be locally Lorentz-invariant while a locally Lorentz-invariant law is not necessarily Lorentz-invariant. It is weaker to demand a law to be locally Lorentz-invariant. The change from the requirement of Lorentz invariance to that of the local Lorentz invariance on laws of physics provides with a larger acceptable scope to explore these laws.

There is a great need for this change. The continuous efforts to construct Lorentz-invariant statistical mechanics and thermodynamics for many-particle systems began soon after Einstein's work on special relativity. But all these efforts have failed. So far we have not had a Lorentz-invariant statistical mechanics and thermodynamics for many-particle systems. In the framework of special relativity, the concepts of sized particles and rigid bodies (systems of particles) are not allowed. All particles are point-like or string-like. We are not entitled to have a Lorentz-invariant mechanics for rigid bodies. We have Lorentz-invariant theories of fields and quantized fields but they, as well as string and superstring theories [4], have suffered from the divergence problem. Phenomenological substituting several finite experimental values of particle masses and charge for their infinities in theoretical calculations, physicists developed renormalization techniques to remove all divergence in some quantized fields. However, not all quantized fields are renormalizable, and it is hard to accept such a kind of renormalizability as a basic physical principle. Moreover, as Feynman said: "renormalization of a quantity gives up any possibility of calculating that quantity" [5], the renormalized theory of quantized fields fails to explain a class of important phenomena, mass differences in the groups of neutron-proton, the $\pi$-mesons, the K-mesons, the $\Sigma$-baryons and the $\Xi$-baryons. It is not futile to expect a locally Lorentz-invariant statistical mechanics and thermodynamics for many-particle systems, a locally Lorentz-invariant mechanics for rigid bodies and a locally Lorentz-invariant quantum field theory with no divergence problem.

There is a hint for this change. According to Einstein's theory of gravitation, the trajectory of an object under gravitational interaction is independent of the mass of the object, so the presence of gravitational forces must be locally equivalent to the use of non-inertial frame of reference. This is the equivalence principle. The word "locally" means a sufficiently small region in space and time. The locality of the equivalence principle lets the local Lorentz invariance, not Lorentz invariance, stand in Einstein's theory of gravitation: In the presence of gravitational field, all non-gravitational laws of physics are locally Lorentz-invariant in the freely falling coordinate system [6].

Experimental facts are in favor of this change. Experiments directly supporting the Lorentz invariance can be divided into several groups relevant to: the constancy of the speed of light, the Einstein time dilation, the Einstein velocity addition law, relativistic mass-velocity and mass-energy relations, and relativistic Doppler effects. About the constancy of the speed of light, two research groups, of Turner and Hill [7], and of Champeney et al [8], placed a $Co^{57}$ source near the rim of a standard centrifuge with an iron absorber near the axis of rotation. They used the Mossbauer effect to look for any velocity dependence of the frequency of the 14.4 KeV $\gamma$-rays as seen by the $Fe^{57}$ in the absorber. They established limits of $\Delta c/c<2 \times 10^{-10}$ for the anisotropy in the one-way speed of light. Riis and his colleagues [9] compared the frequency of a two-photon transition in a fast atomic beam to that of a stationary absorber while the direction of the fast beam is rotated relative to the fixed stars and found the upper limit $\Delta c/c<3.5 \times 10^{-9}$ firstly and $\Delta c/c<2 \times 10^{-11}$ later for the anisotropy. The experiment of Krisher et al [10] was made using highly stable hydrogen-maser frequency standards (clocks) separated by over 21 km and connected by a ultrastable fiber optics link. The limits yielded from the experimental data are respectively $\Delta c/c<2 \times 10^{-7}$ for linear dependency and $\Delta c/c<2 \times 10^{-8}$ for quadratic dependency on the velocity of the Earth with respect to the cosmic microwave background. The Einstein time dilation was verified by Bailey et al [11] to an accuracy of $1 \times 10^{-3}$ and by Kaivola et al [12] to an accuracy of $4 \times 10^{-5}$. More accurate result, $2.3 \times 10^{-6}$, can be found in the report made by R. W. McGowan et al [13]. The Einstein velocity addition law reads

$$y_2^r = \sqrt{1-\frac{u^2}{c^2}} \left\{ (y_1^r - u^r) + \left(\frac{1}{\sqrt{1-\frac{u^2}{c^2}}} - 1\right) u^r \frac{u^s(y_1^s - u^s)}{u^2} \right\} / \left[1 - \frac{y_1^k u^k}{c^2}\right], \quad r,s,k=1,2,3, \tag{4}$$

where $y_m^r = dx_m^r/dt_m$, r=1,2,3, m=1,2. When $y_m^1 = y_m$, $u^1 = u$ and $y_m^2 = y_m^3 = u^2 = u^3 = 0$, it becomes

$$y_2 = (y_1 - u)/[1 - \frac{y_1 u}{c^2}].$$

This velocity addition law is quite consistent with the Fizeau experiment on the light speed in moving liquids [14,15]. In regard to the relativistic mass-velocity and mass-energy relations for a particle,

$$m = m_0/\sqrt{1 - y^2/c^2} \text{ and } E = m_0 c^2/\sqrt{1 - y^2/c^2},$$

Ref.[16] contains a complete discussion on their experimental facts and a long list of these experimental facts. Relativistic Doppler effect is expressed in equation

$$\omega = \gamma \omega_0 (1 - y\cos\phi/c),$$

for a plane electromagnetic wave whose frequency is $\omega_0$ in a laboratory and $\omega$ as observed in a moving inertial frame of reference with relative velocity $y^r$, $y=(y^r y^r)^{1/2}$, where $\phi$ is an angle between wave vector and velocity $y^r$, $\gamma$ is in Eq.(2). It is the transverse Doppler shift when $\phi=\pi/2$, which never exists in the pre-relativistic equation of Doppler effect. The relativistic Doppler effect is in accord with experiments [12,13]. As the local Lorentz invariance involves the constancy of the speed of light, the Einstein velocity addition law and the Einstein time dilation as well as Lorentz invariance, as the local Lorentz invariance leads to the same equations of motion for a particle and for a plane electromagnetic wave as Lorentz invariance [17], as the localized Lorentz transformation can be extended to Lorentz transformation in the case of a plane electromagnetic wave (see below, in Section 5), all experiments directly supporting Lorentz invariance also support the local Lorentz invariance. We do not have any experimental fact which is within the Lorentz invariance but beyond the local Lorentz invariance.

3. DERIVATION OF LORENTZ TRANSFORMATION

Two fundamental postulates stated by Einstein in his derivation of Lorentz transformation are (i) the principle of relativity and (ii) the constancy of the speed of light in all inertial frames of reference. Conceptually, the principle of relativity means that there exists a class of equivalent inertial frames of reference, any one of which moves with a non-zero constant velocity relative to any other and any one of which is supplied with motionless, rigid unit rods of equal length and motionless, synchronized clocks of equal running rate. Einstein wrote: "in a given inertial frame of reference the coordinates mean the results of certain measurements with rigid (motionless) rods, a clock at rest relative to the inertial frame of reference defines a local time, and the local time at all points of space, indicated by synchronized clocks and taken together, give the time of this inertial frame of reference." [18] As defined by Einstein, in each inertial frame of reference, an observer can employ his motionless-rigid rods and motionless-synchronized clocks in the so-called "motionless-rigid rod and motionless-synchronized clock" measurement method to measure space and time intervals. By using this "motionless-rigid rod and motionless-synchronized clock" measurement method, the observer in each inertial frame of reference sets up his usual inertial coordinate system, $\{x^r,t\}$, r=1,2,3. Postulate (ii) asserts that the measured speed of light is the same constant c in every such usual inertial coordinate system.

However, for derivation of the Lorentz transformation, another assumption is necessary besides the two postulates. This other assumption concerns the Euclidean structure of gravity-free space and the homogeneity of gravity-free time in the usual inertial coordinate system,

$$dX^2 = \delta_{rs} dx^r dx^s, \quad r,s=1,2,3, \tag{5a}$$
$$dT^2 = dt^2, \tag{5b}$$

everywhere and every time. Postulates (i) and (ii) and the assumption Eqs.(5a-5b) together yield the Lorentz transformation between any two usual inertial coordinate systems. Indeed though this assumption was not explicitly articulated, evidently having been considered self-evident, Einstein said in 1907: "Since the propagation velocity of light in empty space is c with respect to both reference systems, the two equations, $x_1^2 + y_1^2 + z_1^2 - c^2 t_1^2 = 0$ and $x_2^2 + y_2^2 + z_2^2 - c^2 t_2^2 = 0$ must be equivalent." [3]. Leaving aside the question whether postulate (i) implies the linearity of transformation between any two usual inertial coordinate systems and the reciprocity of relative velocities between any two usual inertial coordinate systems, we know that these two equivalent equations, the linearity of transformation and the reciprocity of relative velocities exactly lead to Lorentz transformation.

Some physicists explicitly articulated the assumption Eqs.(5a-5b) in their works on the topic. Pauli wrote: "This also implies the validity of Euclidean geometry and the homogeneous nature of space and time." [19]. Fock said: "The logical foundation of these methods is, in principle, the hypothesis that Euclidean geometry is applicable to real physical space together with further assumptions, viz. that rigid bodies exist and that light travels in straight lines." [20].

Introducing the four-dimensional usual inertial coordinate system $\{x^\gamma\}$, $\gamma=1,2,3,4$, $x^4=ict$, and the Minkowskian structure of four-dimensional gravity-free spacetime in this coordinate system,

$$d\Sigma^2=\delta_{\alpha\beta}dx^\alpha dx^\beta,\ \alpha,\beta=1,2,3,4,$$

Minkowski [21] showed that Lorentz transformation is just a rotation in this four-dimensional spacetime. The Minkowskian structure is a four-dimensional version of the assumption Eqs.(5a-5b).

## 4. LOCAL STRUCTURES OF GRAVITY-FREE SPACE AND TIME

To derive the localized Lorentz transformation, we keep two postulates (i) and (ii) but renew the assumption Eqs.(5a-5b).

We assume that gravity-free space and time possess the following non-Euclidean structures in the usual inertial coordinate system $\{x^r, t\}$, $r=1,2,3$,

$$dX^2=g_{rs}(dx^1,dx^2,dx^3,dt)dx^r dx^s,\ r,s=1,2,3, \tag{6a}$$
$$dT^2=g(dx^1,dx^2,dx^3,dt)dt^2, \tag{6b}$$
$$g_{rs}(dx^1,dx^2,dx^3,dt)=K^2(y)\delta_{rs}, \tag{6c}$$
$$g(dx^1,dx^2,dx^3,dt)=(1-y^2/c^2)\equiv g(y), \tag{6d}$$
$$K(y)=\frac{c}{2y}(1-y^2/c^2)^{1/2}\ell n\frac{c+y}{c-y}, \tag{6e}$$
$$y=(y^s y^s)^{1/2},\ y<c, \tag{6f}$$
$$y^s=dx^s/dt, \tag{6g}$$

where $dX$ and $dT$ are respectively the real space and time differentials between two neighboring points $(x^1,x^2,x^3,t)$ and $(x^1+dx^1,x^2+dx^2,x^3+dx^3,t+dt)$.

The "motionless-rigid rod and motionless-synchronized clock" measurement method is not all that each inertial frame of reference has. For each inertial frame of reference, we imagine other measurement methods that are different from the "motionless-rigid rod and motionless-synchronized clock" measurement method. By taking these other measurement methods, an observer in each inertial frame of reference can set up other inertial coordinate systems, just as well as he can set up his usual inertial coordinate system by taking the "motionless-rigid rod and motionless-synchronized clock" measurement method. We call these other inertial coordinate systems the unusual inertial coordinate systems. One of the unusual inertial coordinate systems is the primed inertial coordinate system, denoted by $\{x'^r, t'\}$, $r=1,2,3$.

We do believe in flatness of gravity-free space and time. We further assume that gravity-free space and time have the Euclidean structures in the primed inertial coordinate system,

$$dX^2=\delta_{rs}dx'^r dx'^s,\ r,s=1,2,3, \tag{7a}$$
$$dT^2=dt'^2, \tag{7b}$$

everywhere and every time.

The non-Euclidean structures of gravity-free space and time in the usual inertial coordinate system specified by two metric tensors $g_{rs}(dx^1,dx^2,dx^3,dt)$ and $g(dx^1,dx^2,dx^3,dt)$ are of the so-called generalized Finsler geometry [22-26]. The generalized Finsler geometry is a generalization of Riemann geometry. It can be endowed with the Cartan connection [26]. When and only when $y$ approaches zero, metric tensors $g_{rs}(dx^1,dx^2,dx^3,dt)$ and $g(dx^1,dx^2,dx^3,dt)$ become the Euclidean.

## 5. THE LOCALIZED LORENTZ TRANSFORMATION

Two assumptions Eqs.(6a-6g) and Eqs.(7a-7b) and two postulates (i) and (ii) together yield the localized Lorentz transformation between any two usual inertial coordinate systems.

Introducing $y'^s=dx'^s/dt'$, $s=1,2,3$, we have from Eqs.(6a-6g) and Eqs.(7a-7b),

$$y'^r=[\frac{c}{2y}\ell n\frac{c+y}{c-y}]y^r,\ r=1,2,3, \tag{8a}$$

and
$$y' = \frac{c}{2} \ln \frac{c+y}{c-y}, \qquad (8b)$$
where $y'=(y'^s y'^s)^{1/2}$, $s=1,2,3$. We name $y'^s$ the primed velocity. Eqs.(8a-8b) specify the relationship between a Newtonian velocity $y^r$ and its corresponding primed velocity $y'^r$. In Eq.(8b), as $y$ goes to $c$, we find infinite primed speed,
$$c' = \lim_{y \to c} \frac{c}{2} \ln \frac{c+y}{c-y}.$$
Primed speed $c'$ is actually a new version of the speed of light, the speed of light in the primed inertial coordinate system. On the invariant Newtonian speed $c$ in all usual inertial coordinate systems, primed speed $c'$ is invariant in all primed inertial coordinate systems.

Let $\{x^r_m, t_m\}$ and $\{x'^r_m, t'_m\}$ be respectively the usual and the primed inertial coordinate systems of IFR m, m=1,2, where IFR2 moves with non-zero Newtonian velocity $u^s$, s=1,2,3, relative to IFR1. Since of the same $c'$ in both $\{x'^r_1, t'_1\}$ and $\{x'^r_2, t'_2\}$ and of the assumption Eqs.(7a-7b), two equations
$$\delta_{rs} dx'^r_1 dx'^s_1 - c'^2 (dt'_1)^2 = 0, \qquad (9a)$$
$$\delta_{rs} dx'^r_2 dx'^s_2 - c'^2 (dt'_2)^2 = 0, \qquad (9b)$$
or two equations
$$\delta_{rs} x'^r_1 x'^s_1 - c'^2 (t'_1)^2 = 0, \qquad (10a)$$
$$\delta_{rs} x'^r_2 x'^s_2 - c'^2 (t'_2)^2 = 0, \qquad (10b)$$
must be equivalent for the propagation of light, everywhere and every time. For light propagation, Eqs.(6a-6g) carry $y \to c$. Using Eqs.(7a-7b) and Eqs.(6a-6g) with $y \to c$ in Eqs.(9a-9b), we find other two equivalent equations,
$$\delta_{rs} dx^r_1 dx^s_1 - c^2 (dt_1)^2 = 0, \qquad (11a)$$
$$\delta_{rs} dx^r_2 dx^s_2 - c^2 (dt_2)^2 = 0, \qquad (11b)$$
everywhere and every time, because $c^2 K^2(c) = c'^2 g(c)$, where $K(c) = \lim_{y \to c} K(y)$, $g(c) = \lim_{y \to c} g(y)$.

Two equivalent equations Eqs.(10a-10b), the linearity of transformation between two $\{x'^r_m, t'_m\}$ and the reciprocity of relative velocities between two $\{x'^r_m, t'_m\}$ lead to the $c'$-type Galilean transformation between two primed inertial coordinate systems $\{x'^r_m, t'_m\}$, m=1,2. Two equivalent equations Eqs.(11a-11b), the linearity of transformation between two $\{x^r_m, t_m\}$, and the reciprocity of relative velocities between two $\{x^r_m, t_m\}$ lead to the localized Lorentz transformation, Eqs.(3a-3b), between two usual inertial coordinate systems $\{x^r_m, t_m\}$, m=1,2.

In general, gravity-free space and time have different non-uniform structures in two different usual inertial coordinate systems $\{x^r_m, t_m\}$, m=1,2. The localized Lorentz transformation between these two usual inertial coordinate systems can not be extended to Lorentz transformation, in other words, we have no longer the Lorentz transformation between them. In the case that gravity-free space and time possess identical uniform structures in these two usual inertial coordinate systems, the localized Lorentz transformation between them can be extended to the Lorentz transformation. A plane electromagnetic wave in empty space (y=c) is of the case.

6. RELATIVISTIC VELOCITY SPACE

Let us look at some direct consequences of the assumptions Eqs.(6a-6g) and Eqs.(7a-7b). Dividing Eq.(6a) by Eq.(6b), we find
$$Y^2 = [\frac{c}{2y} \ln \frac{c+y}{c-y}]^2 \delta_{rs} y^r y^s, \quad r,s=1,2,3. \qquad (12)$$
With the calculation techniques in Riemann geometry, one can prove that Eq.(12) embodies what the equations
$$dY^2 = H_{rs}(y) dy^r dy^s, \quad r,s=1,2,3, \qquad (13a)$$
$$H_{rs}(y) = c^2 \delta^{rs}/(c^2-y^2) + c^2 y^r y^s/(c^2-y^2)^2, \text{ real } y^r \text{ and } y<c, \qquad (13b)$$
do. Similarly, dividing Eq.(7a) by Eq.(7b), we have
$$Y^2 = \delta_{rs} y'^r y'^s, \quad r,s=1,2,3, \qquad (14a)$$

which embodies what the equation
$$dY^2=\delta_{rs}dy'^r dy'^s, \quad r,s=1,2,3, \tag{14b}$$
does. Eqs.(13a-13b) and (14b) represent the same velocity space that is respectively defined in the usual velocity-coordinates $\{y^r\}$ and the primed velocity-coordinates $\{y'^r\}$, r=1,2,3. We call this velocity space the relativistic velocity space. The relativistic velocity space is characterized by a finite boundary at y=c and the Einstein velocity addition law in the usual velocity-coordinates and by unboundedness and the Galilean velocity addition law in the primed velocity-coordinates [17]. Eqs.(13a-13b) and Eq.(14b) imply
$$dy'^r = A^r_s(y)dy^s, \quad r,s=1,2,3, \tag{15a}$$
$$A^r_s(y)=\gamma\delta^{rs}+\gamma(\gamma-1)y^r y^s/y^2 \tag{15b}$$
because
$$\delta_{rs}A^r_p(y)A^s_q(y)=H_{pq}(y), \quad p,q=1,2,3.$$

The geometric structures of gravity-free space and time in the usual and the primed inertial coordinate systems respectively match that of the relativistic velocity space in the usual and the primed velocity-coordinates. Generating two representations of the relativistic velocity space in the usual and the primed velocity-coordinates, such matching enables us to get the relativistic generalization of Maxwell's velocity distribution [27-29].

The Euclidean structure of the relativistic velocity space in the primed velocity-coordinates convinces us of Maxwell's distribution of primed velocities,

$$P(y'^1,y'^2,y'^3)dy'^1 dy'^2 dy'^3 = N\left(\frac{m}{2\pi K_B T}\right)^{3/2} \exp\left[-\frac{m}{2K_B T}(y')^2\right] dy'^1 dy'^2 dy'^3, \tag{16a}$$

$$P(y')dy' = 4\pi N\left(\frac{m}{2\pi K_B T}\right)^{3/2} (y')^2 \exp\left[-\frac{m}{2K_B T}(y')^2\right] dy'. \tag{16b}$$

Inserting Eqs.(15a-15b) and (8b) in Eqs.(16a-16b), we obtain the relativistic equilibrium distribution of Newtonian velocities,

$$P(y^1,y^2,y^3)dy^1 dy^2 dy^3 = N\frac{(m/2\pi K_B T)^{3/2}}{(1-y^2/c^2)^2} \exp\left[-\frac{mc^2}{8K_B T}\left(\ell n \frac{c+y}{c-y}\right)^2\right] dy^1 dy^2 dy^3, \tag{17a}$$

$$P(y)dy = \pi c^2 N\frac{(m/2\pi K_B T)^{3/2}}{(1-y^2/c^2)} \left(\ell n \frac{c+y}{c-y}\right)^2 \exp\left[-\frac{mc^2}{8K_B T}\left(\ell n \frac{c+y}{c-y}\right)^2\right] dy. \tag{17b}$$

The relativistic equilibrium velocity distribution fits to the Maxwellian distribution for low-energy particles (y<<c) but substantially differs from the Maxwellian distribution for high-energy particles. It falls off to zero as y goes to c.

The relativistic equilibrium velocity distribution has been used to explain the observed non-Maxwellian decay mode of high-energy tails in velocity distributions of astrophysical plasma (planetary magnetospheres, solar wind and other) particles [30]. The deviation of the decay mode of these high-energy tails from Maxwellian has been observed for many years [31-34]. Experimental data were mostly, if not all, modeled by the $\kappa$ (kappa) distribution. As experimental data seem to be well modeled by the kappa distribution, as the kappa distribution contains a power-law decay when y goes to infinity, it was concluded that the decay mode of high-energy tails in velocity distributions of astrophysical plasma particles is power-law like. This conclusion is rather misleading. The kappa distribution shapes

$$K(y)dy = \frac{N}{\pi^{3/2}} \frac{1}{\theta^3} \frac{\Gamma(\kappa+1)}{\kappa^{3/2}\Gamma(\kappa-1/2)} \left(1+\frac{y^2}{\kappa\theta^2}\right)^{-(\kappa+1)} dy \tag{18}$$

where $\theta = [(2\kappa-3)/\kappa]^{1/2}(K_B T/m)^{1/2}$, $\Gamma$ is the gamma function and kappa is a parameter to be determined in comparison with experimental data [33,34]. Different values of kappa correspond to different kinds of velocity distribution. When and only when kappa goes to infinity, the kappa distribution becomes the Maxwellian. The kappa distribution can not be a good modeling distribution for experimental data. The reasons are: For any value of kappa, the kappa distribution extends as far as y=$\infty$, while experimental data, we believe, as far as y=c; Velocity distribution of low-energy particles, as observed, can be well described by the Maxwellian distribution, but the kappa distribution with any finite kappa value does not reduce to the Maxwellian even for small y; The values of kappa in fitting experimental data vary from event to event [32]. The relativistic equilibrium velocity distribution predicts a new decay mode for those high-energy tails: falling off to zero, as y goes to c, slower than any exponential decay, $\exp\{-[2c/(c-y)]^B\}$, and faster than any power-law decay, $(c-y)^n$, where B and n are two positive numbers [30].

The relativistic equilibrium velocity distribution has been also used to calculate the nuclear fusion reaction rate [35,36]. To create a nuclear fusion reaction, a proton or nucleus must penetrate the repulsive Coulomb barrier and be close to another proton or nucleus so that the strong interaction between them acts. The Coulomb barrier is in general much higher than thermal energy, so nuclear fusion reactions can occur only among few high-energy protons and nuclei. If, under the conditions for nuclear fusion reactions, interacting protons and nuclei reach their equilibrium distribution in the period of time that is infinitesimal compared to the mean lifetime of nuclear fusion reactions, it is the equilibrium velocity distribution of these few high-energy protons and nuclei that participates in determining the rate of nuclear fusion reactions. In this circumstance, it is inappropriate to use the Maxwellian velocity distribution to calculate the nuclear fusion reaction rate [36]. We have to use the relativistic equilibrium velocity distribution for the purpose. The calculation results indicate that the nuclear fusion reaction rate based on the relativistic equilibrium distribution, R, has a reduction factor with respect to that based on the Maxwellian velocity distribution, $R_M$,

$$R = \frac{\tanh Q}{Q} R_M, \tag{19a}$$

$$Q = (2\pi z_1 z_2 \frac{K_B T}{\mu c^2} \frac{e^2}{\hbar c})^{1/3}, \tag{19b}$$

where the reduction factor, $\tanh Q/Q$, depends on temperature T, reduced mass $\mu$ and atomic numbers $z_1$ and $z_2$ of the studied nuclear fusion reactions. In other words, the reduction factor varies with the kind of neutrinos. Since $0 < Q < \infty$, the reduction factor takes values between 0 and 1, $0 < \tanh Q/Q < 1$, that gives rise to

$$0 < R < R_M. \tag{20}$$

Eqs.(19a-19b) and (20) signify much in resolving the solar neutrino problem. The relativistic equilibrium velocity distribution is a possible solution to the solar neutrino problem [36-39].

To match the gravity-free space and time and the relativistic velocity space in their geometric structures is important. As being generalized to the case where gravitational field presents, it leads to a prediction on the velocity distribution of low-energy particles in the presence of spherically symmetric gravitational field, as a test of Einstein's theory of gravitation [40].

7. CONCLUSION AND DISCUSSION

(1) Introducing the primed inertial coordinate system, for each inertial frame of reference, in addition to the usual inertial coordinate system, we have assumed that gravity-free space and time possess the Euclidean structures in the primed inertial coordinate system and the generalized Finslerian structures in the usual inertial coordinate system.

(2) We have combined these two assumptions with two fundamental postulates, (i) the principle of relativity and (ii) the constancy of the speed of light in all inertial frames of reference, to derive the localized Lorentz transformation as a linear transformation between any two usual inertial coordinate systems. The theory founded by the two assumptions and the two postulates is called the modified special relativity theory [17,27,41].

(3) It is based on the principle of relativity and the localized Lorentz transformation between any two usual inertial coordinate systems to conclude that all laws of physics are locally Lorentz-invariant in the usual inertial coordinate system.

(4) As the Lorentz transformation covers the localized Lorentz transformation and the localized Lorentz transformation does not necessarily lead to Lorentz transformation, a Lorentz-invariant law of physics must be locally Lorentz-invariant while a locally Lorentz-invariant law is not necessarily Lorentz-invariant. The change from the requirement of Lorentz invariance to that of the local Lorentz invariance on laws of physics provides with a larger acceptable scope to explore these laws.

(5) It is the "motionless-rigid rod and motionless-synchronized clock" measurement method that we use in our experiments. All our experimental data are collected and expressed in the usual inertial coordinate system. To get locally Lorentz-invariant laws of physics in the usual inertial coordinate system, we can take the physical principle of local Lorentz invariance: All physical laws keep their forms under the localized Lorentz transformation. This physical principle must be incorporated into the generalized Finslerian structures of gravity-free space and time in the usual inertial coordinate system. However, we have an alternative physical principle for getting locally Lorentz-invariant laws of physics in the usual inertial coordinate system [17].

(6) According to our recent work [17], the locally Lorentz-invariant mechanics for a particle is the same as Lorentz-invariant one and the locally Lorentz-invariant quantum field theory is divergence-free.

(7) The localization of Lorentz transformation and its induced local Lorentz invariance precisely meet the demands of Einstein's theory of gravitation.

(8) Our assumptions on the local structures of gravity-free space and time are experimentally verifiable.

In this paper, we did not define the primed inertial coordinate system from the measurement point of view. We did not discuss the measurement contents involved in the primed velocity, either. We prefer doing these somewhere else [42].

ACKNOWLEDGMENT
The author greatly appreciates the teachings of Prof. Wo-Te Shen. The author thanks Prof. S. S. Bandola and Dr. P. Rucker for helpful suggestions.

REFERENCES
[1]     H. A. Lorentz, Proc. Acad. Sc., 6, 809 (1904)
[2]     A. Einstein, Ann. Physik, 17, 891 (1905)
[3]     A. Einstein, Jarbuch der Radioaktivitat und Elektronik, 4, 411 (1907), reprinted in The Collected Papers of A. Einstein, vol.2, 252, Princeton University Press, Princeton, NJ (1989)
[4]     Nathan Berkovits, hep-th/9707242
[5]     R. P. Feynman, The present status of quantum electrodynamics, in The Quantum Theory of Fields, ed. R. Stops, Interscience Publishing Co., New York (1962)
[6]     C. M. Will, gr-qc/0103036
[7]     K. C. Turner and H. A. Hill, Phys. Rev., 134, B252 (1964)
[8]     D. C. Champeney, G. R. Isaak and A. M. Khan, Phys. Lett., 7, 241 (1963)
[9]     E. Riis et al, Phys. Rev. Lett., 60, 81(1988)
[10]    T. P. Krisher et al, Phys. Rev., D45, 731(1990)
[11]    J. Bailey et al, Nature (London), 268, 301 (1977)
[12]    M. Kaivola et al, Phys. Rev. Lett., 54, 255 (1985)
[13]    R. W. McGowan et al, Phys. Rev. Lett., 71, 251 (1993)
[14]    P. Zeeman, Arch. Neerl. Sci. Exaxtes. Nat., 3A10, 131 (1927)
[15]    I. Lerche, Am. J. Phys., 45, 1154 (1977)
[16]    Special Relativity Theory, selected papers, ed. , American Institute of Physics, New York (1962)
[17]    Jian-Miin Liu, Chaos Solitons&Fractals, 12. 1111 (2001)
[18]    A. Einstein, Autobiographical Notes, in: A. Einstein: Philospheo-Scientist, ed. P. A. Schipp, 3rd edition, Tudor, New York (1970)
[19]    W. Pauli, Theory of Relativity, Pergamon Press Ltd., New York (1958), translator G. Field
[20]    V. Fock, The Theory of Space Time and Gravitation, Pergamon Press, New York (1959)
[21]    H. Minkowski, Raum und Zeit, Phys. Z., 10, 104-111 (1909)
[22]    P. Finsler, Uber Kurven und Flachen in Allgemeinen Raumen, Dissertation, Gottingen (1918), Birkhauser Verlag, Basel (1951)
[23]    E. Cartan, Les Espaces de Finsler, Actualites 79, Hermann, Paris (1934)
[24]    H. Rund, The Differential Geometry of Finsler Spaces, Springer-Verlag, Berlin (1959)
[25]    G. S. Asanov, Finsler Geometry, Relativity and Gauge Theories, D. Reidel Publishing Company, Dordrecht (1985)
[26]    Jian-Miin Liu, physics/0208047
[27]    Jian-Miin Liu, Chaos Solitons&Fractals, 12, 2149 (2001)

[28]    Jian-Miin Liu, cond-mat/0108356
[29]    Jian-Miin Liu, cond-mat/0301042
[30]    Jian-Miin Liu, cond-mat/0112084
[31]    M. P. Leubner, J. Geophys. Res., 105, 27387 (2000)
[32]    S. P. Christon et al, J. Geophys. Res., 93, 2562 (1988)
[33]    V. M. Vasyliunas, J. Geophys. Res., 73, 2839 (1968)
[34]    D. Summers and R. M. Thorne, Phys. Fluids, B3, 1835 (1991)
[35]    Jian-Miin Liu, nucl-th/0210058
[36]    Jian-Miin Liu, Relativistic equilibrium velocity distribution, nuclear fusion reaction rate and the solar neutrino problem, to be posted on astro-ph/0307xxx
[37]    J. N. Bahcall, astro-ph/0209080
[38]    S. Johansson, The solar FAQ: Solar neutrino and other solar oddities, http://www.talkorigins.org
[39]    Jian-Miin Liu, physics/0110002
[40]    Jian-Miin Liu, gr-qc/0206047
[41]    Jian-Miin Liu, Chaos Solitons&Fractals, 12, 399 (2001)
[42]    Jian-Miin Liu, The Modified Special Relativity Theory, to be published